\documentclass[app,amsmath,amssymb,reprint,superscriptaddress,floatfix]{revtex4-1}

\usepackage{graphicx}
\usepackage{dcolumn}
\usepackage{bm}

\usepackage[utf8]{inputenc}
\usepackage[T1]{fontenc}
\usepackage{mathptmx}
\usepackage{siunitx}
\usepackage{hyperref}

\begin{document}


\title{High-Frequency GaAs Optomechanical Bullseye Resonator}

\author{N. C. Carvalho}
\email{natccar@ifi.unicamp.br}
\affiliation{Applied Physics Department and Photonics Research Center, University of Campinas, Campinas, SP, Brazil}
\author{R. Benevides}
\affiliation{Applied Physics Department and Photonics Research Center, University of Campinas, Campinas, SP, Brazil}
\author{M. M\'{e}nard}
\affiliation{Department of Computer Science, Universit\'{e} du Qu\'{e}bec \`{a} Montr\'{e}al, Montr\'{e}al, Canada}
\author{G. S. Wiederhecker}
\affiliation{Applied Physics Department and Photonics Research Center, University of Campinas, Campinas, SP, Brazil}
\author{N. C. Frateschi}
\affiliation{Applied Physics Department and Photonics Research Center, University of Campinas, Campinas, SP, Brazil}
\author{T. P. Mayer Alegre}
\email{alegre@unicamp.br}
\affiliation{Applied Physics Department and Photonics Research Center, University of Campinas, Campinas, SP, Brazil}


\begin{abstract}
The integration of optomechanics and optoelectronics in a single device opens new possibilities for developing information technologies and exploring fundamental phenomena. Gallium arsenide (GaAs) is a well-known material that can bridge the gap between the functionalities of optomechanical devices and optical gain media. Here, we experimentally demonstrate a high-frequency GaAs optomechanical resonator with a ring-type bullseye geometry that is unprecedented in this platform. We measured mechanical modes up to $\SI{3.4}{\GHz}$ with quality factors of 4000 (at $\SI{77}{\kelvin}$) and optomechanical coupling rates up to $\SI{39}{\kHz}$ at telecom wavelengths. Moreover, we investigated the material symmetry break due to elastic anisotropy and its impact on the mechanical mode spectrum. Finally, we assessed the temperature dependence of the mechanical losses and demonstrated the efficiency and anisotropy resilience of the bullseye anchor loss suppression, indicating that lower temperature operation may allow mechanical quality factors over $10^4$. Such characteristics are valuable for active optomechanics, coherent microwave-to-optics conversion via piezo-mechanics and other implementations of high-frequency oscillators in III-V materials.
\end{abstract}

\maketitle

\section*{\label{sec:level1}Introduction}

The engineering of light-matter interaction in optomechanical devices has allowed the observation of very relevant fundamental phenomena such as gravitational waves~\cite{Abbott2016} and ground-state cooling~\cite{Chan2011}; and consequently has enabled important developments in information science technology~\cite{Hill2012, OConnell2010}. For the past decade, silicon has been the material of choice for most on-chip optomechanics experiments. However, as we move towards high power efficiency and quantum-level control, device design becomes more challenging and material properties more restrictive, which have driven intense research into alternative materials~\cite{Guha2017b,morozov2019,zou2016, mitchell2019}. Therefore, gallium arsenide (GaAs) arises as a mature platform with the potential to match or overcome silicon in many properties, such as light confinement and optomechanical coupling strength, due to its high-refractive-index and large photo-elastic coefficients~\cite{balram2014}. Moreover, the optical losses that often impair the performance of GaAs have been mitigated by improved etching and surface passivation techniques, leading to optical quality factors ($Q_\text{opt}$) of over a million~\cite{Guha2017a}.

Developing optomechanical resonators based on III-V materials would not only allow for their disruptive integration with coherent light sources and single quantum emitters, but also open routes to explore the interplay of gain and loss in non-Hermitian physical systems~\cite{Ozdemir2019}. Besides its advantageous optoelectronic properties, GaAs also has other valuable characteristics, such as piezoelectricity, revealing its suitability for wavelength conversion mediated by piezo-optomechanics~\cite{Wu2020}, enabling thus the ultimate integration and control of charge carriers, light, sound, and microwave fields.

To unleash the outstanding properties of GaAs and enable an optomechanical device operation in the resolved-sideband regime, a design supporting high mechanical frequencies must be devised. In simple microdisks, this is limited by the resonator radius, which turns to be unpractically small \cite{baker2014, ding2010, ding2011}. Also, it becomes relatively complex in optomechanical crystals, which requires complicated designs \cite{balram2014}. Despite recent efforts with nanobeam cavities showing impressive optomechanical coupling rates, mechanical frequencies have not exceeded $\SI{2.8}{\GHz}$ in GaAs~\cite{Balram2016, Forsch2020}. In this work, we demonstrate GaAs optomechanical devices built using a bullseye design that allows for very high mechanical quality factors and measured mechanical modes up to $\SI{3.4}{\GHz}$, with the potential to explore even higher frequency modes, according to simulations. Our design allows for optimizing the optical and mechanical resonances independently, offering to couple between several high-quality factors optical modes with a single mechanical mode, which can be interesting for the exploration of multi-mode optomechanical experiments.

\begin{figure*}[!ht]
\includegraphics[width=1\textwidth]{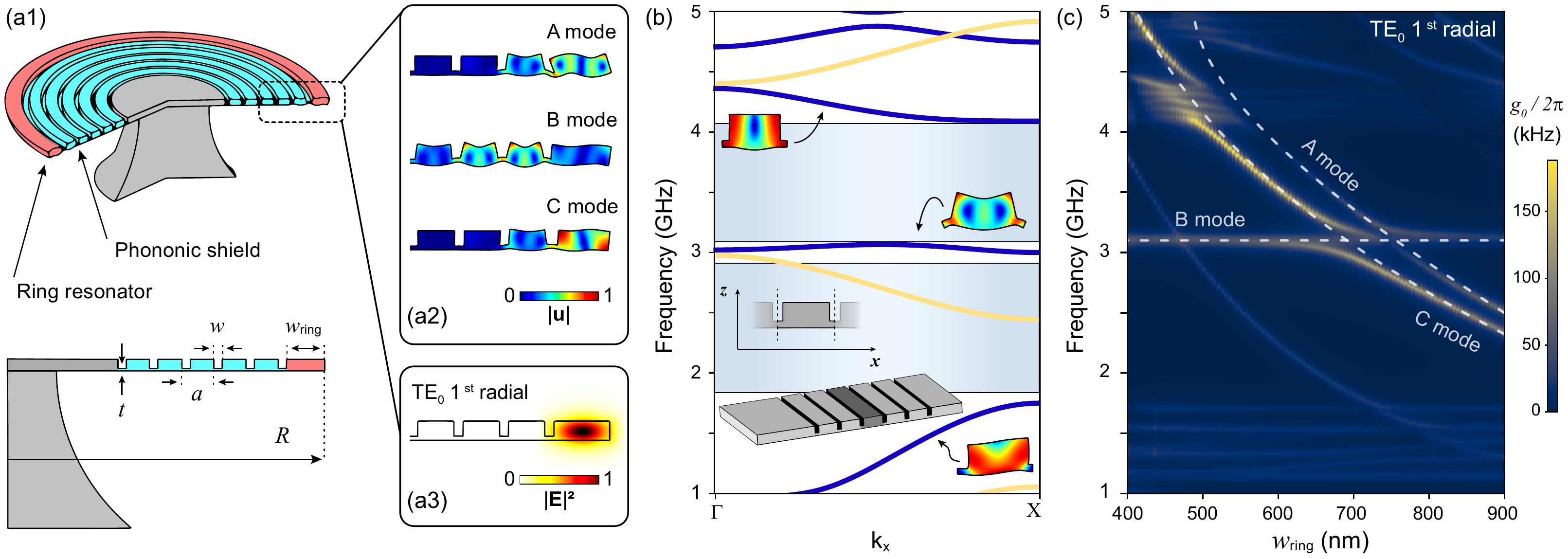}
\caption{\label{fig1}~(a1) Illustration of the bullseye resonator (top) and geometric parameters (bottom). Inset: (a2) simulated displacement profile of the mechanical modes and (a3) electric amplitude of the whispering gallery optical mode with transverse electric polarization and fundamental radial-order (bottom). (b) The blue (yellow) lines are the mechanical bands of the bullseye grating simulated as a linear crystal for x-polarized or z-polarized (y-polarized) modes. The shaded blue show the target bandgap and the insets contain the mechanical deformation for the modes at the band-edge (X-point). (c) 2D simulation results of the optomechanical coupling rate ($g_0/(2\pi)$), the gray dashed lines are drawn as a guide to the eyes. The grating parameters in (b) and (c) are $a= \SI{600}{\nm}$, $w= \SI{120}{\nm}$ and $t = \SI{50}{\nm}$.}
\end{figure*}

Our samples were fabricated using a technique that allows different etching depths to be obtained in a single-step lithography process. The mechanical performance of our device was investigated by cooling it to liquid nitrogen temperatures and by characterizing typical sources of mechanical dissipation, revealing that the mechanical quality factor is only limited by temperature-dependent effects, which can be circumvented at cryogenic conditions. We also show that our GaAs samples are strongly affected by the material elastic anisotropy and numerically evaluate its implications, obtaining a good match between experimental data and simulations. The devices achieved optomechanical coupling rates and mechanical quality factors up to 39 kHz and 4000 (at $\sim \SI{80}{K}$), respectively, operating just above the resolved sideband limit.

\section*{\label{sec:level2}Design and fabrication}

The bullseye geometry, originally devised by Santos \textit{et al.}~\cite{Santos2017} using silicon, consists of a ring-type cavity resonator obtained by patterning a nano-structured circular grating over a micro-disk, as shown in the diagram in Fig.~\ref{fig1}~(a1). In this design, the mechanical waves are confined to the outermost ring using a radial phononic shield, which also isolates the ring cavity from the supporting pedestal, inhibiting anchor losses and, therefore, enabling very high mechanical quality factors. A key advantage of the bullseye design, compared to nanobeams or optomechanical crystal devices, is the complete decoupling between optical and mechanical resonance frequencies. The former is mostly influenced by the radius of the disk, $R$ in Fig.~\ref{fig1}~(a1), whereas the latter will be defined by the external ring width ($w_\mathrm{ring}$).
In this way, mechanical frequencies can be increased by narrowing down the external ring, with minimal impact on the optical frequencies, as long as $w_\mathrm{ring}\gtrsim\SI{500}{\mu m}$ ~\cite{Santos2017}. 

Our GaAs bullseye project was based on a nominal geometry with a disk radius of $\SI{6}{\um}$ and grating dimensions set to $a=\SI{600}{\nm}$, $w=\SI{120}{\nm}$ and $t=\SI{50}{\nm}$, as shown in Fig.~\ref{fig1}~(a1). This grating is designed as a phononic shield to confine dilatational mechanical waves propagating in the radial direction and, as long as their wavelengths are small compared to the ring radius, one can neglect its curvature and approximate it to a linear crystal \cite{Santos2017} (Fig.~\ref{fig1}~(b) - inset). Using 2D cartesian Finite Element Method (FEM), we modeled a linear grating with the geometrical parameters mentioned above, obtaining a band-diagram as in Fig.~\ref{fig1}~(b). Such approximation revealed a partial phononic bandgap for longitudinal ($x$-polarized) and vertical shear waves ($z$-polarized), represented by the blue lines in Fig.~\ref{fig1}~(b), where shear-horizontal waves ($y$-polarized) are also shown in yellow. It resulted in two mechanical frequency stopbands in the ranges between $\sim \SI{2}{\GHz}~-~\SI{3}{\GHz}$ and $\sim\SI{3}{\GHz}~-~\SI{4}{\GHz}$, able to constrain radial dilatational mechanical waves to the edge of the disk.

The optomechanical coupling rate, $g_0/(2\pi)$, for the full bullseye structure is shown in Fig.~\ref{fig1}~(c). Our simulations (2D - axisymmetric) show that three mechanical modes couple to the (whispering gallery-type) optical field, as highlighted by the gray dashed lines between $\SI{2}{}$ and $\SI{5}{\GHz}$. The mechanical displacement profiles of these modes are shown in Fig.~\ref{fig1}~(a2), followed by the first radial-order transverse electric (TE) optical mode to which they are coupled (in Fig.~\ref{fig1}~(a3)). It reveals that C is the desired ring-type breathing mode, while A is a flexural ring mode, and B is a grating mode that is almost independent of the external ring size. Unlike GaAs disk resonators with similar dimensions, the $g_0$ for the first-order ring dilatational mechanical mode is dominated by the photo-elastic effect; precisely, our calculations for a \SI{740}{nm} $w_\mathrm{ring}$ show that the moving boundaries \cite{johnson2002} and photo-elastic \cite{chan2012} optomechanical coupling rates are \SI{12}{kHz} and \SI{129}{kHz}, respectively, resulting in a total $g_0/(2\pi)$ of \SI{141}{kHz}.

\begin{figure}[hbt!]
\includegraphics [width=1\linewidth]{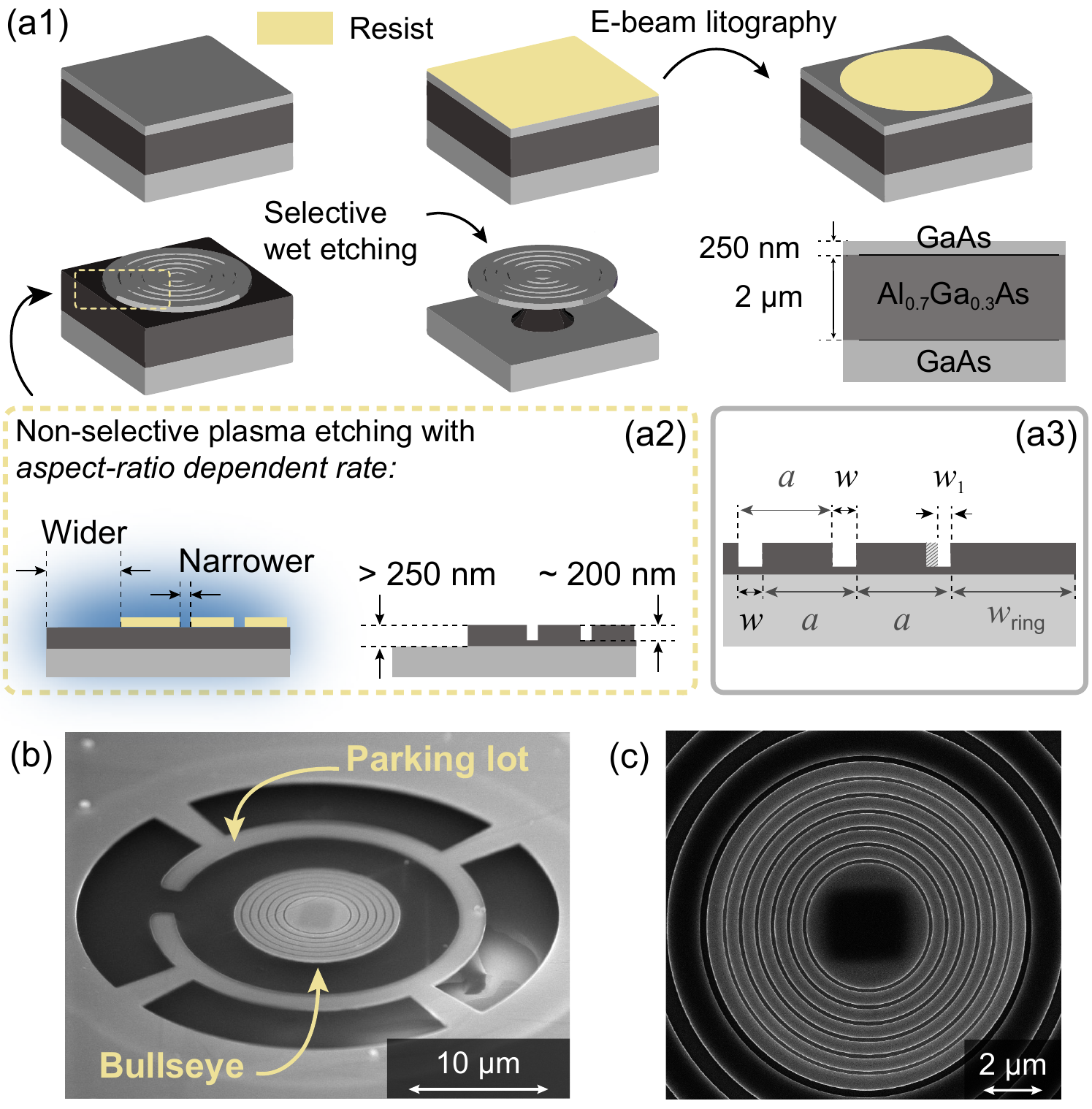}
\caption{\label{fig2}(a1) Simplified illustration of the bullseye fabrication steps. (a2) and (a3) contain the diagrams high-lightening the plasma etching rate contrast for regions of different sizes, and the narrowing of the first outer groove due to the electron beam proximity effect, respectively. (b) Scanning electron microscope image of a typical device (bullseye resonator surrounded by the \textit{parking lot}, a structure designed to stabilize the tapered fiber loop). (c) Top-view of a sample.}
\end{figure}

The devices were fabricated from a GaAs/Al$_{0.7}$Ga$_{0.3}$As stack ($\SI{250}{\nm}/\SI{2000}{\nm}$) that was grown over a GaAs substrate using molecular beam epitaxy (Canadian Photonics Fabrication Centre). In contrast with the previous silicon bullseye, which was fabricated with an optical stepper, the GaAs structure was defined using electron beam lithography~\cite{Benevides2020}. The fabrication steps of the GaAs bullseye samples are summarized in Fig.~\ref{fig2} (a1). The devices were patterned on the top of the GaAs wafer through a single plasma etching step. We used positive electro-resist (ZEP-520A) and managed to define grating grooves and remove all the material outside the microdisk region using the aspect-ratio dependent etching, where the etching rate of narrower gaps is lower in comparison to wider regions (Fig.~\ref{fig2} (a2)) \cite{shaqfeh1989,bailey1995}. The bullseye disks were then released from the substrate by selective wet etching of the AlGaAs buffer layer with hydrofluoric acid (HF), followed by standard organic cleaning. This advantageous technique avoided the need for multiple lithography steps that would introduce complex alignment procedures. 

Figures \ref{fig2} (b) and (c) present scanning microscope images of exemplar samples. The former shows the final device surrounded by the \textit{parking lot}, which was designed to support the fiber loop. Top-view images, such as in Fig.~\ref{fig2} (c), were used to characterize the dimensions of the devices. The disk radius ($R$), $w_\mathrm{ring}$, $a$ and $w$ were measured through this method, whereas $t$ was only estimated from the etching rate, which was calibrated before the fabrication of the devices. Electron beam proximity effects caused the first outer groove ($w_1$ in Fig.~2~(a3)) to have a narrower width in comparison to the inner grooves ($w$). Nevertheless, we obtained very regular and sharp grooves that are only slightly angled. Section S1, in the Supplementary Material, discusses in detail the impact of the verticality of the sidewalls and the depth of the grooves in the mechanical grating structure. Finally, in section S2, both $w_1$ and $t$ were found by matching the measured and simulated values for the optical mode dispersion of our device. All those geometrical values were then used in the FEM simulations of Fig.~\ref{fig3}.

\section*{\label{sec:level3}Results and discussion}

\begin{figure*}[ht!]
\includegraphics[width=1\textwidth]{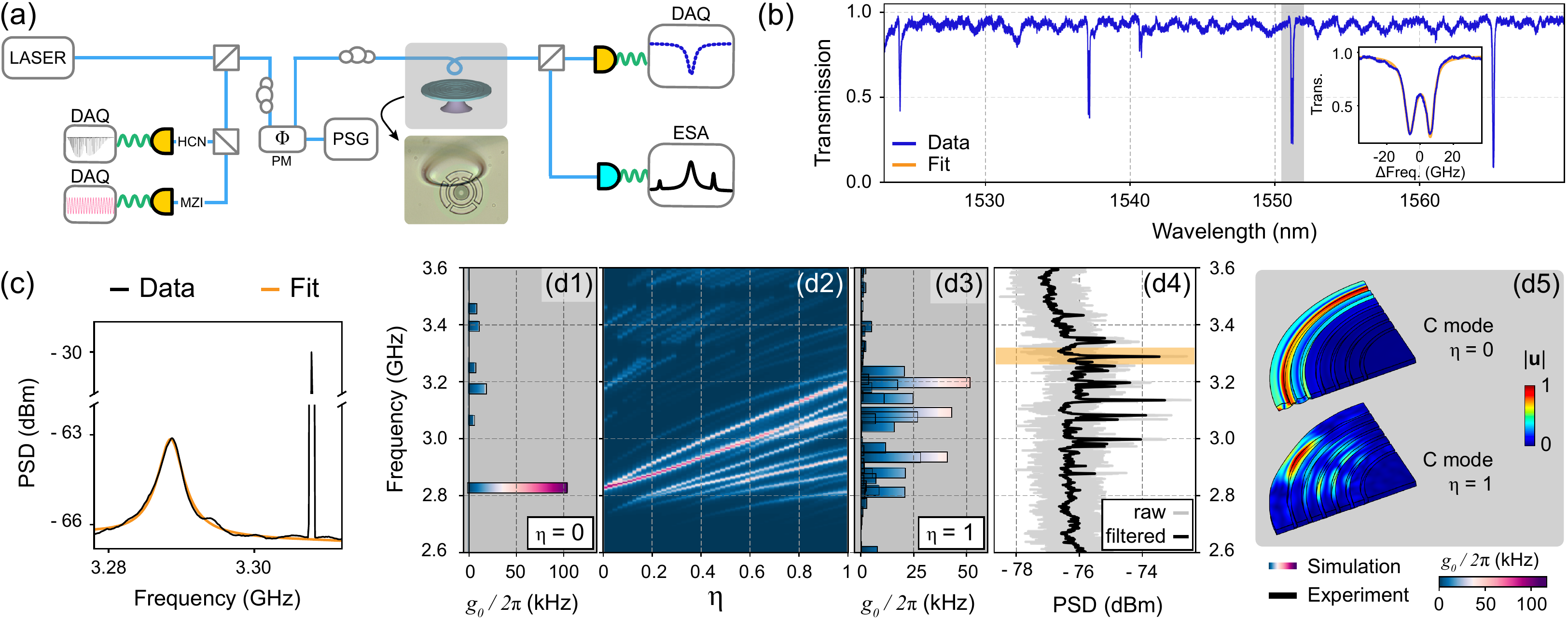}
\caption{\label{fig3}~(a) Optomechanical characterization setup (DAQ - Data Acquisition System, PSG - Power Signal Generator, ESA - Electrical Spectrum Analyzer, PM - Phase Modulator, HCN - Hydrogen Cyanide wavelength reference and MZI - Mach–Zehnder Interferometer). Inset: optical microscope image of the tapered fiber loop coupled to a device. (b) Optical transmission spectrum of the $w_\mathrm{ring}
~=~\SI{740}{\nm}$ sample. The inset shows the optical modes (whispering gallery-type of transverse electric polarization and second radial-order) highlighted in light gray. The orange line is the Lorentzian fit. (c) Power spectral density (PSD) of the selected mechanical mode of the $\SI{740}{\nm}$ sample (highlighted in shaded orange in (d4)) with respective $g_0$ calibration peak (the orange line is the Lorentzian fit).  3D FEM simulations of: (d1) the mechanical modes when $\eta = 0$ (fully isotropic), (d2) the optomechanical coupling rate as a function of the material anisotropy parameter ($\eta$) and (d3) mechanical modes when $\eta = 1$ (fully anisotropic). (d4) Broad experimental PSD (raw and digital filtered data using the Savitzky–Golay method \cite{SG}). (d5) Simulated mechanical displacement profiles of the 2.8 GHz and the 3.2 GHz modes (highest bars) in (d1) and (d3), respectively. The information in (d1)-(d5) corresponds to the $w_\mathrm{ring}
~=~\SI{740}{\nm}$ and all experimental data were acquired at room temperature and atmospheric pressure.}
\end{figure*}

The optomechanical properties of the resonators were characterized in the setup schematically represented in Fig.~\ref{fig3}~(a). The bullseye microcavity was probed via evanescent coupling with a tapered fiber loop continuously fed by a tunable C-band laser (New Focus TLB-6728). The transmitted optical signal was then split and simultaneously measured by a slow and a fast detector. The DC component gave information about the optical response of the cavity, while the fast signal was measured by an electrical spectrum analyzer that read out the mechanical mode signatures imparted on the transmitted optical signal.

We experimentally investigated two devices with different $w_\mathrm{ring}$ sizes: $\SI{700}{\nm}$ and $\SI{740}{\nm}$, at room temperature and atmospheric pressure. The optical spectrum of the $\SI{740}{\nm}$ sample is displayed in Fig.~\ref{fig3}~(b). The inset has a mode doublet (counter-propagating whispering gallery modes) presenting total linewidths, $\kappa/(2\pi)$, of $\sim\SI{6}{\GHz}$ ($Q_{\text{opt}} \sim \SI{3e4}{}$), which was a typical value measured in our samples. These modes provided the strongest readout of the mechanical modes and were found to be second radial-order modes with quasi-TE polarization (major component along the radial direction). The identification was based on the comparison between the measured and simulated frequency dependence of the free spectral range (due to group velocity dispersion)~\cite{fujii2020} - see Supplementary Material, S2, for details.

The optomechanical coupling with the second radial-order TE optical mode is very similar to its fundamental counterpart (Fig. \ref{fig1} (c)) and predicts $g_0/(2\pi)= \SI{136}{kHz}$ for the sample with a ring size of 740 nm. The complete 2D simulation for this optical mode structure as a function of the $w_\mathrm{ring}$ size can be found in the Supplementary Material, S2. Although a slightly larger $g_0$ is expected for the optomechanical coupling through the first radial-order mode, our measurement scheme was sensitive to $g_0^2/\kappa$ and, therefore, the detection of the mechanical interaction with such optical modes was compromised by their larger linewidths.

A 3.29 GHz mechanical resonance was measured in the $\SI{740}{\nm}$ $w_\mathrm{ring}$ and shown in Fig.~\ref{fig3}~(c). The Lorentzian fit gave a mechanical quality factor ($Q_{\text{m}}$) of 900 and the calibration of $g_0$ --~through the comparison between the phase modulator RF-tone and the optical transduction of the cavity~\cite{Gorodetksy2010}~-- resulted in an optomechanical coupling rate of $g_0/(2\pi)=\SI{34}{\kHz}$. Analogously, we obtained a $Q_{\text{m}}$ of 1200 and a $g_0/(2\pi)=\SI{39}{\kHz}$ for a $\SI{3.13}{\GHz}$ mode of the $\SI{700}{\nm}$ $w_\mathrm{ring}$ sample. Numerical simulations of $g_0$ for the $\SI{740}{\nm}$ $w_\mathrm{ring}$ are shown in Fig.~\ref{fig3}~(d1)-(d3). Also, Fig.~\ref{fig3}~(d4) presents a $\SI{1}{\GHz}$-broad experimental mechanical spectrum - raw and Savitzky–Golay filtered~\cite{SG}. The data used to obtain the optomechanical coupling rates was not filtered.

The lower $g_0$ and the multiple peaks observed in the experimental spectra of Fig.~\ref{fig3}~(d4) are not consistent with axisymmetric simulations that predict a single peak for each mechanical mode (A, B and C in Fig.~\ref{fig1}~(a1)). Therefore, we employed a three-dimensional FEM model to account for the well-known elastic anisotropy of GaAs and precisely identify the mechanical modes. Indeed, as the material anisotropy is gradually increased in the 3D simulations, a clear degeneracy lifting of the mechanical mode frequencies is observed, as shown in Fig.~\ref{fig3}~(d1)-(d3), by sweeping the anisotropy parameter $\eta$. Here, $\eta = 0$ corresponds to an isotropic device and $\eta = 1$ is the full anisotropic case, according to the relation $c^*_{44}(\eta)=\eta c_{44}+(1-\eta)((c_{11}+c_{22})/2)$ for the GaAs stiffness tensor component.

\begin{figure}[ht!]
\includegraphics [width=0.98\linewidth]{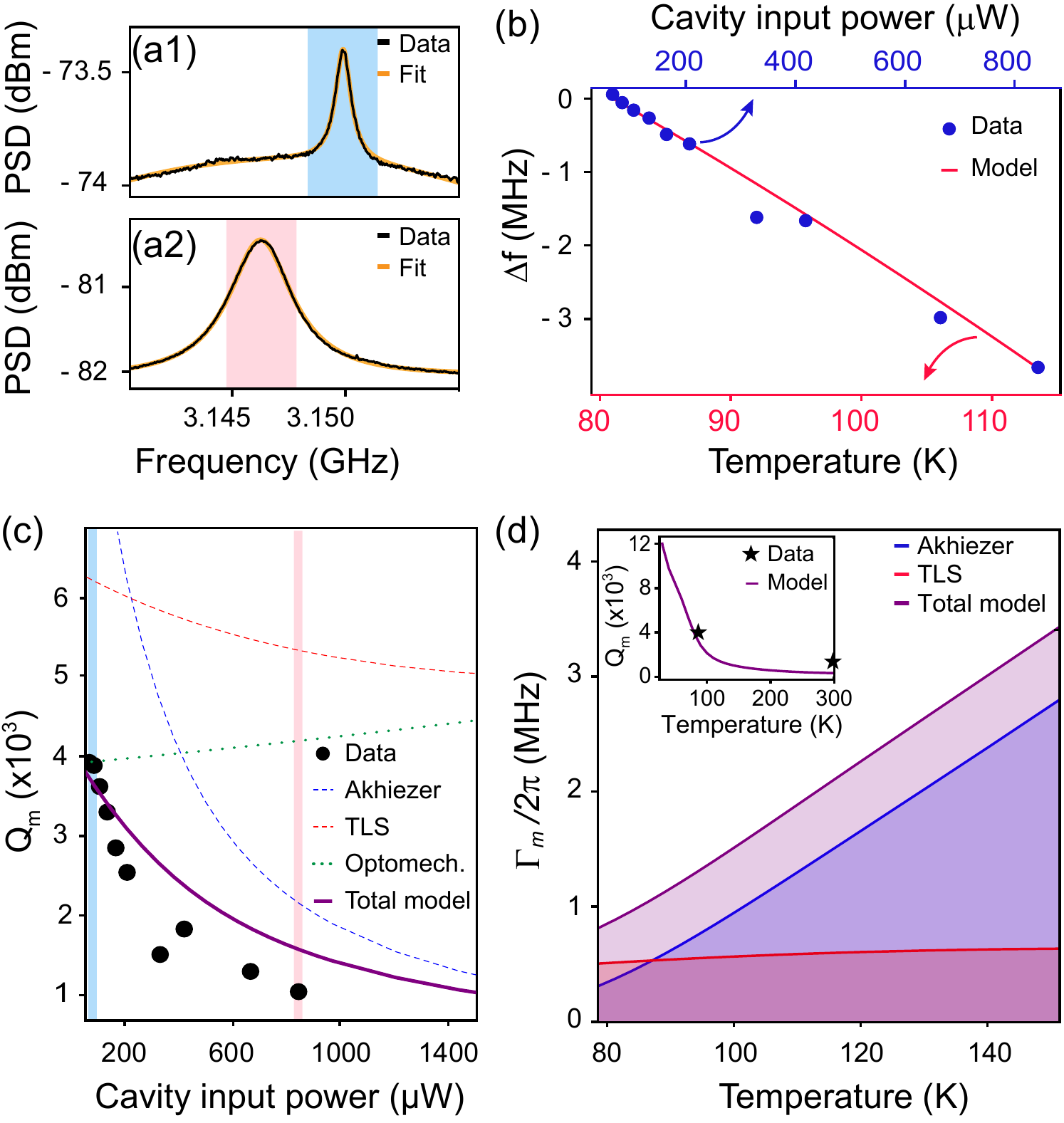}
\caption{\label{fig4}~Temperature dependence of the $\SI{3.1}{\GHz}$ mode of the $\SI{700}{\nm}$ ring width sample: PSD of the mechanical resonance measured for the lower (light blue - (a1)) and higher (light red - (a2)) cavity input power (Lorentzian fits in orange); (b) measured mechanical frequency shift ($\Delta f$) as a function of the cavity power (blue) and computed frequency shift resulting from the $w_{\text{ring}}$ thermal expansion as a function of temperature (red). (c) Comparison of the experimental mechanical Q-factor ($Q_{\text{m}}$) power dependence to the total model of $Q_{\text{m}}$ (Akhiezer, TLS and optomechanical anti-damping) and the calculated optomechanical backaction effect only (narrowing of the mechanical linewidth with increasing power). Akhiezer and TLS individual contributions are shown. Light blue and red strips correspond to the data of the resonant modes shown in (a1) and (a2), respectively. (d) Nanomechanical dissipation model of the bullseye resonator. Inset: calculated $Q_{\text{m}}$ compared to the experimental data of low and room temperature measurements.}
\end{figure}

The isotropic mechanical displacement profile shown in Fig.~\ref{fig3}~(d5) ($\eta = 0$) corresponds to the prominent bar in Fig.~\ref{fig3}~(d1). It demonstrates that only the C-mode is optomechanically coupled to the second radial-order optical mode. Indeed, despite the modification of the mode structure after the loss of symmetry in the material, the measured mechanical mode, highlighted in shaded orange in Fig.~\ref{fig3}~(d4), can still be related to the C-mode (external ring breathing mode) as can be seen in Fig.~\ref{fig3}~(d5) ($\eta = 1$), which corresponds to the highest bar in (d3). The 3D modeling also shows that the mechanical grating isolation is resilient to the anisotropy, but predicts a reduced $g_0$ when compared to the 2D simulations with isotropic material elasticity. Thus, when $\eta = 1$, it foresees a $g_0/(2\pi)$ of up to \SI{53}{kHz} (\SI{33}{kHz} due photo-elastic effect and \SI{20}{kHz} associated to the moving boundaries contribution - see section S2, in the Supplementary Material), which agrees reasonably well with the calibrated data presented in Fig.~\ref{fig3}~(c). Moreover, such result is still above the achievable $g_0$ for the fundamental radial breathing mechanical mode in GaAs disks with the same radius and thickness ($g_0^{\text{(disk)}}/(2\pi) \sim (\omega_0/R)x_{\text{zpf}} = \SI{12}{kHz}$, where $\omega_0$ is the angular optical frequency and $x_{\text{zpf}}$ is the zero-point amplitude fluctuation of a 230 MHz mode), with the benefit of having mechanical GHz frequencies not yet reported in any GaAs optomechanical structure operating at telecom wavelengths.

In order to investigate the role of the mechanical grating in inhibiting clamping losses, thermal channels of dissipation had to be suppressed. Therefore, our samples were cooled down to the temperature of liquid nitrogen. Figure~\ref{fig4}~(a1) contains the measured mechanical C-mode of the $\SI{700}{\nm}$ $w_\mathrm{ring}$. At a temperature of $\sim$ \SI{80}{\K}, a $Q_\text{m}$ of $4\times10^3$ was measured, which is an improvement of over three times in comparison to room temperature data. To observe the behavior of the mechanical linewidth ($\Gamma_\text{m}$) as a function of the cavity input power, we performed a laser power sweep - frequency tuned for maximum optomechanical transduction, on the blue side of the optical resonance. As the power was increased, we noticed a simultaneous increase in the mechanical linewidth and a red-shift, $\Delta f = f - f_0$, where $f_0$ is the initial mechanical mode frequency and $f$ is its frequency at a given optical input power, as shown in Fig.~\ref{fig4}~(a2).

In Fig.~\ref{fig4}~(b) we show that $\Delta f$ goes down a few $\SI{}{\MHz}$ when approaching $\SI{1}{\mW}$ of optical input power. This effect is explained by the cavity heating, which causes the material to expand, modifying the mode frequency~\cite{benevides2018}. We calculated this shift and plotted it against the data by assuming that at very low power this heating was negligible, i. e., $\Delta f = 0$, and that the initial reading of the temperature sensor was a good approximation for the cavity temperature. Then, it was possible to estimate a linear relationship between temperature and the input power to the resonator, which was done by comparing the measured power induced frequency shift to the expected shift caused by thermal expansion. 

We also measured the decrease of $Q_{\text{m}}$ with power, as displayed in Fig.~\ref{fig4}~(c), that contains the data of Fig.~\ref{fig4}~(a1) and (a2) in the extreme input power values, highlighted in light blue (low power) and light red (high power), respectively. In order to understand this behavior, we included other loss mechanisms to the mechanical linewidth. When neglecting anchor losses, the linewidth broadening of nanomechanical resonators is then in general dominated by phonon-phonon interactions and scattering by defects. The former can introduce losses via relaxation of thermal phonons, whereas the second dissipates mechanical energy by coupling strain waves to two-level systems (TLS)~\cite{Mohanty2002, Phillips1987}. 

To estimate the linewidth of our sample, we calculated the anharmonic and the TLS mechanical attenuation through the methods described in Ref.~\cite{Vacher2005} (see S3 in the Supplementary Material). 
From \SI{300}{\kelvin} to \SI{80}{\kelvin}, our bullseye resonators falls into the Akhiezer regime \cite{Alegre2011}, $\omega_m \tau_{ph} \lesssim 1$, where $\tau_{ph}$ is the phonon relaxation time for this channel and $\omega_m$ is the mechanical resonator angular frequency. Thermoelastic losses obtained from FEM were found to be several orders of magnitude lower than the Akhiezer damping and are neglected in this analysis. The double-well potentials parameters of our TLS dissipation model were obtained from Ref. \cite{Hamoumi2018}, where GaAs microdisks were investigated; as such values are highly material dependent and do not change significantly with small variation in geometry, they serve as a reasonable approximation for our devices. The results are displayed in Fig.~\ref{fig4}~(d), where the inset includes a direct comparison of the measured mechanical quality factors ($\SI{3.1}{\GHz}$ mode) to the theory, which predicts that $Q_{\text{m}} \sim 12 \times 10^3$ at 20 K. It is important to notice that, at very low temperatures, scattering by defects is the main dissipation mechanism, thus predictions may not be so accurate in this regime. 

The measurement of $Q_{\text{m}}$ as the input laser power was increased gives a hint about the accuracy of our mechanical dissipation model, which correctly predicts the $Q_{\text{m}}$ optical power dependence, as shown in Fig.~\ref{fig4}~(c). The total model not only accounted for the thermal sources of mechanical dissipation (Akhiezer and TLS) but also included the linewidth modification expected from optomechanical backaction. If the measured $Q_{\text{m}}$ was dominated by optomechanical dynamics, an opposite trend should be observed, with increasing (decreasing) Q-factor ($\Gamma_{\text{m}}$) for higher input power. Therefore, we can see that optomechanical anti-damping is not playing a significant role in the mechanical linewidth, indicating that it is dominated by temperature-dependent dissipation. Moreover, the 3D calculation including mechanical clamping and perfectly matched layers predicted mechanical Q-factors between $10^4-10^7$ for the ring modes (Fig.~\ref{fig3}~(d4)), which are at least one order of magnitude higher than the measured values, thus suggesting that our device is not limited by anchor losses.

\section*{\label{sec:level4}Conclusion}

In summary, we designed a GaAs bullseye resonator with a phononic grating operating between $\SI{3}{\GHz}$ and $\SI{4}{\GHz}$. The fabricated devices show modes ranging between $\sim \SI{3.0}{\GHz} - \SI{3.4}{\GHz}$ with optomechanical coupling rates of up to $\SI{39}{\kHz}$ and mechanical quality factor of $\SI{4000}{}$ when cooled down to the temperature of liquid nitrogen, rendering an optomechanical cooperativity of $\text{C}_{\text{om}}\sim 0.05$ at $\SI{80}{\K}$ ($\SI{0.8}{\mW}$ of cavity input power). In order to harness the full potential of the GaAs bullseye resonators, the regime of high cooperativity must be accessible and, thus,  higher optical and mechanical Q-factors must be achieved. The former could be obtained by minimizing roughness~\cite{Benevides2020} and surface absorption~\cite{Guha2017a} with electroresist thermal reflow and alumina passivation ($Q_\text{opt}~\sim~10^5-10^6$). Additionally, mechanical dissipation is expected to be drastically reduced at lower temperatures \cite{Forsch2020} as thermal anharmonic losses are reduced ($Q_\text{m}~\sim~14~\times~10^3$). Under the above conditions, unitary $\text{C}_{\text{om}}$ is reachable with the same $\SI{0.8}{\mW}$ or less power. Incorporating III-V  quantum emitters to the bullseye cavity would also enable the study of active optomechanics, opening a plethora of possibilities, including the creation of an alternative approach for hybrid systems that couples single emitters to mechanical strain~\cite{Yeo2014} and the realization of mechanically modulated light sources~\cite{Princepe2018}. Microwave-to-optical conversion, on the other hand, could take advantage of higher mechanical frequencies enabled by narrower $w_\mathrm{ring}$'s or exploring higher-order mechanical modes. Finally, the bullseye design was shown to be robust across different material platforms and could be extended to other semiconductor materials with lower non-linear optical losses, such as GaP~\cite{Ghorbel2019a, stockill2019}.

\section*{\label{sec:level5} ACKNOWLEDGMENTS}

The authors would like to acknowledge CCSNano-UNICAMP for providing the micro-fabrication infrastructure and CMC Microsystems for providing access to MBE epitaxy and the GaAs wafers. 
This work was supported by S\~{a}o Paulo Research Foundation (FAPESP) through grants 2017/19770-1, 2016/18308-0, 2018/15580-6, 2018/15577-5, 2018/25339-4, Coordena{\c c}\~ao de Aperfei{\c c}oamento de Pessoal de N{\'i}vel Superior - Brasil (CAPES) (Finance Code 001), Financiadora de Estudos e Projetos and the National Sciences and Engineering Research Council (NSERC) of Canada.

\section*{\label{sec:level6} SUPPLEMENTARY MATERIAL}

See Supplementary Material for details on the fabrication of the devices (S1), the characterization of the optical modes through the frequency dependence of the free spectral range (S2) and modeling of the mechanical dissipation mechanisms (S3). 

\section*{\label{sec:level7} DATA AVAILABILITY}
The data that support the findings of this study are available from the corresponding author upon reasonable request.


%

\end{document}



\title{Supplementary Material: High-Frequency GaAs Optomechanical Bullseye Resonator}

\author{N. C. Carvalho}
\email{natccar@ifi.unicamp.br}
\affiliation{Applied Physics Department and Photonics Research Center, University of Campinas, Campinas, SP, Brazil}
\author{R. Benevides}
\affiliation{Applied Physics Department and Photonics Research Center, University of Campinas, Campinas, SP, Brazil}
\author{M. M\'{e}nard}
\affiliation{Department of Computer Science, Universit\'{e} du Qu\'{e}bec \`{a} Montr\'{e}al, Montr\'{e}al, Canada}
\author{G. S. Wiederhecker}
\affiliation{Applied Physics Department and Photonics Research Center, University of Campinas, Campinas, SP, Brazil}
\author{N. C. Frateschi}
\affiliation{Applied Physics Department and Photonics Research Center, University of Campinas, Campinas, SP, Brazil}
\author{T. P. Mayer Alegre}
\email{alegre@unicamp.br}
\affiliation{Applied Physics Department and Photonics Research Center, University of Campinas, Campinas, SP, Brazil}

\maketitle

\section{Fabrication of the GaAs Bullseye Resonator\label{S1}}

The bullseye resonators were obtained through electrolithography with a soft mask etched in an inductively coupled plasma (ICP) with an argon-chlorine mixture. The fabrication steps were based on the recipe proposed in \cite{benevides2018}, but without the thermal reflow of the resist, which would damage the very narrow grooves of the circular grating. Despite careful calibration of the exposure dose, the resolution of the electron beam was still affected by proximity effects, causing the first outer groove ($w_1$) to be narrower than the inner grooves ($w$), as illustrated in Fig.~2~(a3) in the main text. Although other grating parameters, such as $w$ and $a$, were characterized using top images of the samples, $w_1$ was found by matching the measured and simulated values for the optical mode dispersion of our device (see next Section \ref{S2}). 
 
\begin{figure*}[h]
\includegraphics [width=\linewidth]{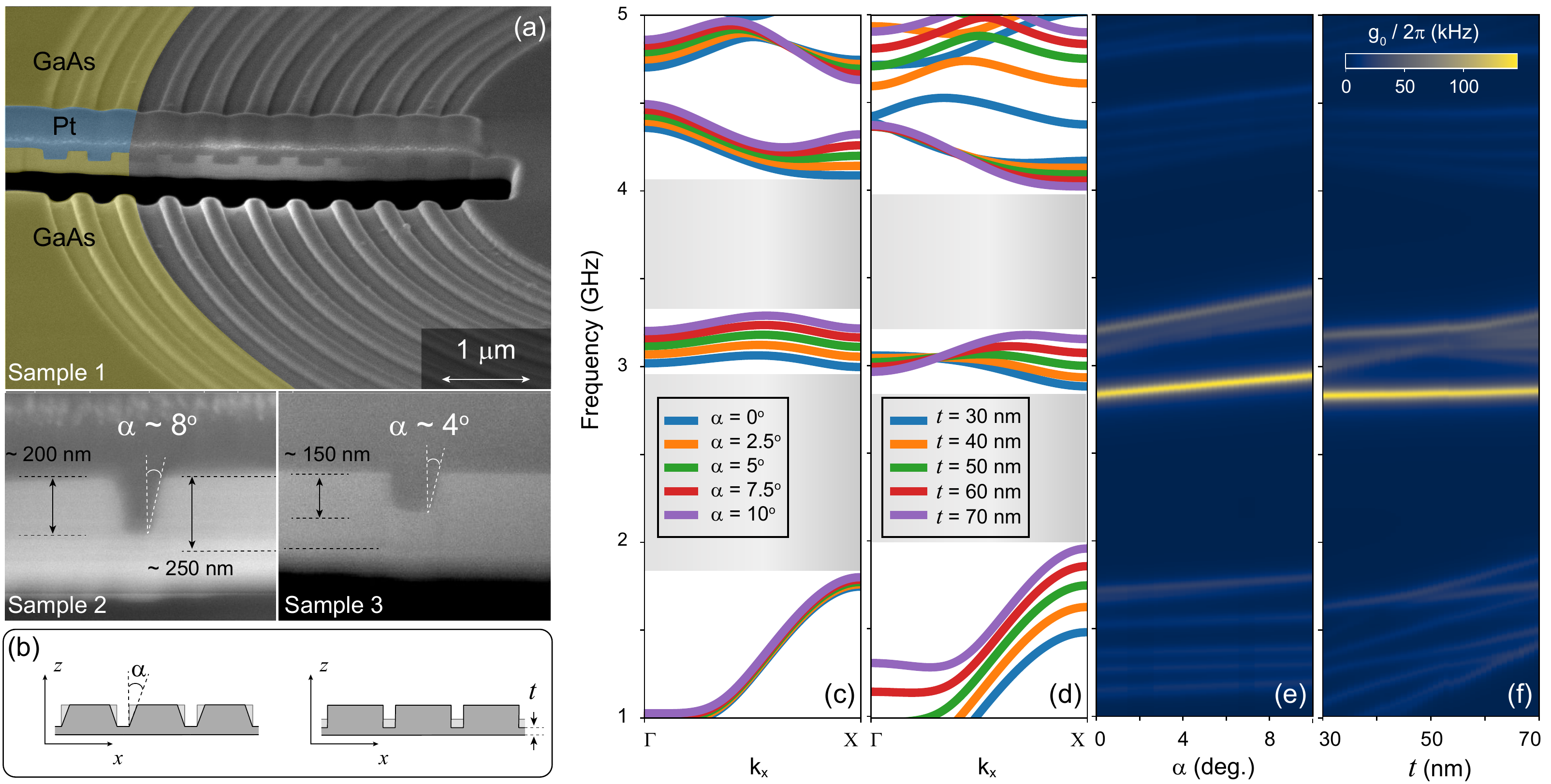}
\caption{(a) Scanning electron microscope images of Samples 1, 2 and 3 - platinum (blue) was deposited over the GaAs samples (yellow) to allow a clean cross-section cut. (b) Illustration of the angled sidewalls and depth of the grooves. (c) and (d): Band-diagrams of the bullseye grating simulated as a linear crystal for x-polarized or z-polarized modes (partial bandgaps are highlighted in light gray). (c) Shows the mechanical bands for different sidewall angles and (d) shows their behavior as a function of $t$. (e) and (f) are 2D FEM simulations of the optomechanical coupling rate ($g_0/(2\pi)$) of the complete bullseye ($w_{\text{ring}} = \SI{740}{nm}$) as a function of $\alpha$ and $t$, respectively. \label{s1}}
\end{figure*}

As explained in the main text, we were able to process the disks and gratings in a single lithography session (electron beam exposure followed by ICP etching) by controlling the aspect ratio dependent etching rate \cite{shaqfeh1989,bailey1995}. To achieve the designed groove depth of \SI{200}{nm}, test samples were characterized to determine the necessary etching time for the chosen dimensions. Fig. \ref{s1} (a) shows Scanning Electron Microscope (SEM) images of three samples, where the cross-sectional cuts were obtained by depositing platinum (Pt) over the GaAs layer (as highlighted in blue and yellow in the image of Sample 1). It is worth noting that Samples 2 and 3 have $w$'s with approximately the designed dimensions ($\sim \SI{120}{nm}$) but present angled walls (deviating by $\alpha$ from the vertical direction). 

The final devices were fabricated by reproducing the recipe of Sample 2. However, fine-tuning of $t$ and $\alpha$ was not possible. Therefore, in Fig. \ref{s1} we also show a Finite Element Method (FEM) analysis of the resilience of the (partial) phononic bandgap ((c) and (d) - 2D cartesian model of the correspondent linear grating) and of the optomechanical performance ((e) and (f) - 2D axisymmetric model of the complete bullseye structure) for variations of $\alpha$ and $t$ near the values measured in the Samples 2 and 3. Such investigation allowed us to conclude that the design of the device is robust against such small deviations. Yet, we decided to also treat $t$ as a free parameter in the optical dispersion analysis as discussed in Section \ref{S2}.

\section{Frequency Dependence of the Free Spectral Range \label{S2}}

In order to determine the spatial distribution and polarization of the measured optical modes, we analyzed the group velocity dispersion of the optical spectra by investigating the frequency dependence of the Free Spectral Range (FSR). Adjacent longitudinal optical modes $\omega_{\mu}$ in relation to a reference mode, $\omega_0$ ($=2\pi\nu_0$), and $\mu$ ($=m-m_0$) can be expanded using the FSR as~\cite{fujii2020}:
\begin{equation} \label{eq1}
\omega_{\mu} = \omega_0 + \mu D_1 + \frac{1}{2}\mu^2 D_2 + \frac{1}{6}\mu^3 D_3 + ... ,
\end{equation}

\noindent where $D_1/2\pi$ is the FSR and $D_2/2\pi$ correspond to its variation rate with $\mu$, and so forth. Notice that in this case, a positive $D_2$ corresponds to anomalous group-velocity dispersion (GVD) for the $\mu$-mode.

\begin{figure*}[h]
\includegraphics [width=\linewidth]{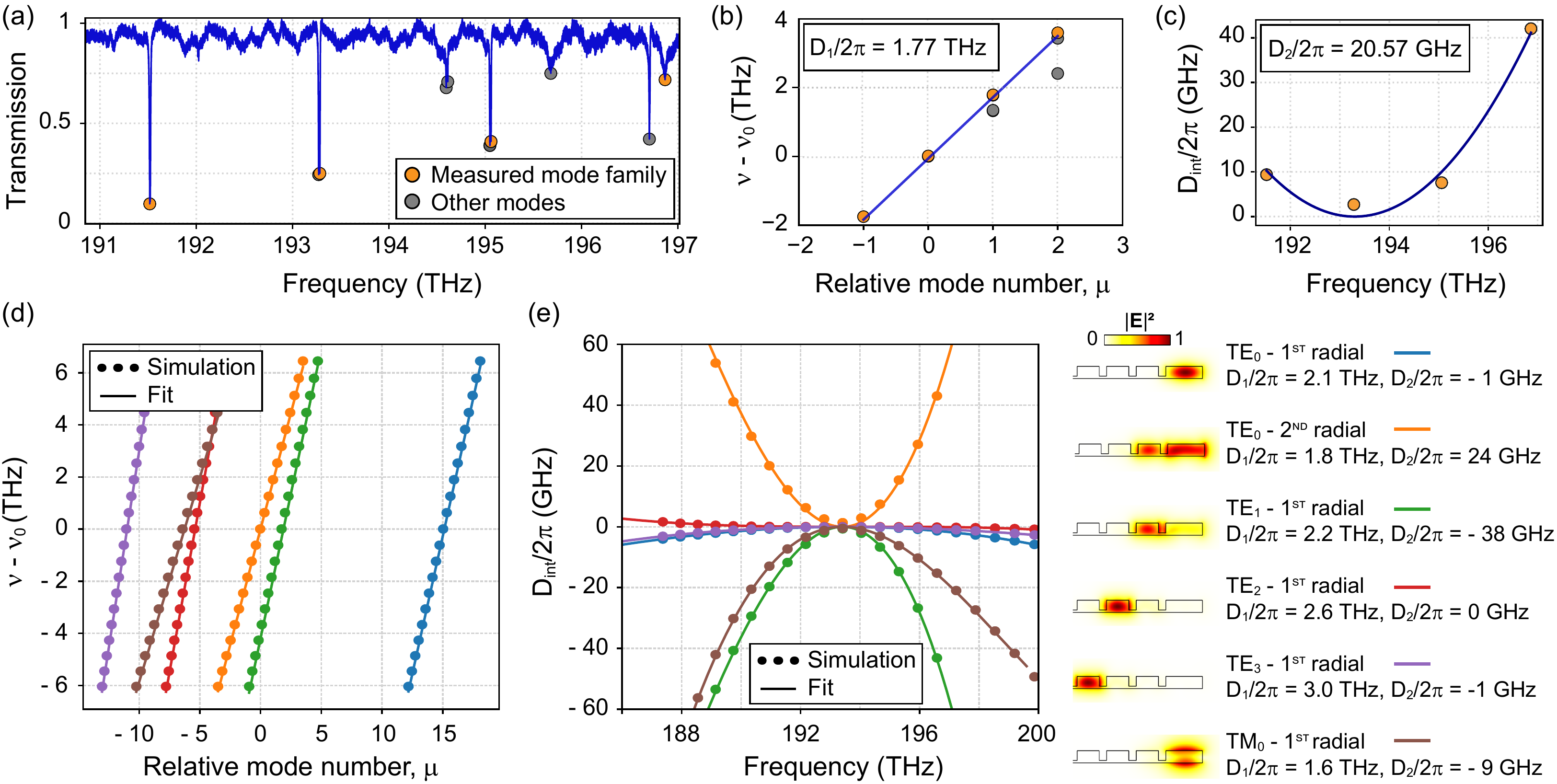}
\caption{(a) Optical spectrum of the $\SI{740}{\nm}$ $w_\mathrm{ring}$ sample highlighting the measured mode family. (b) Linear fitting of the data points ($\nu$ is the mode frequency and $\nu_0 = $ 193.3 THz). (c) Quadratic fitting of the residual dispersion ($D_{\text{int}}$). FEM simulation of the linear (d) and quadratic (e) coefficients. Right panel: model results and respective simulated optical profiles.\label{s2}}
\end{figure*}

In Fig.~\ref{s2} (a) the measured optical spectrum of the $\SI{740}{\nm}$ $w_\mathrm{ring}$ sample is provided. The optical mode at $\SI{193.3}{\THz}$ is used for measuring the mechanical breathing mode on our devices. This mode belongs to a family with a FSR of $\sim$ \SI{1.77}{THz}, as displayed in Fig.~\ref{s2} (b). To emphasize the anomalous dispersion nature for our mode we plot the residual dispersion, $D_{\text{int}} = \omega_{\mu} - \omega_0 - \mu D_1$, in Fig.~\ref{s2} (c). By fitting Eq.~\ref{eq1} to the shifted frequency data shown in Fig.~\ref{s2} (b) a  $D_2/(2\pi) = \SI{20.6}{GHz}$ is found. 

In order to fully identify the optical mode, we used a FEM model to simulate the optical dispersion of the bullseye device that includes both geometrical and material dispersion~\cite{Kachare_1976}. The dimensions used in the simulation were $R = \SI{6}{\micro\metre}$, $w_\mathrm{ring} = \SI{746}{\nm}$, $a = \SI{600}{\nm}$, and $w = \SI{175}{\nm}$, which were recovered from a SEM image as shown in Fig. 2 of the main text. The first outer groove, $w_1$, and the remaining device thickness, $t$, could not be precisely recovered from the SEM. They were therefore treated as free parameters -- within the fabrication uncertainty --  in order to match the simulated $D_1$ and $D_2$ values to the measured ones. Figs.~\ref{s2} (d) and (e) show the FEM model results of our structure for the first six different mode families. The optical mode profile $|E^2|$ and the fitting results for each mode family are also shown. The optical dispersion curves were obtained for $w_1 = \SI{150}{\nm}$ and $t = \SI{50}{nm}$ that, along with the above-mentioned parameters, were assumed to be the dimensions of the real device.  

\begin{figure*}[h]
\includegraphics [width=\linewidth]{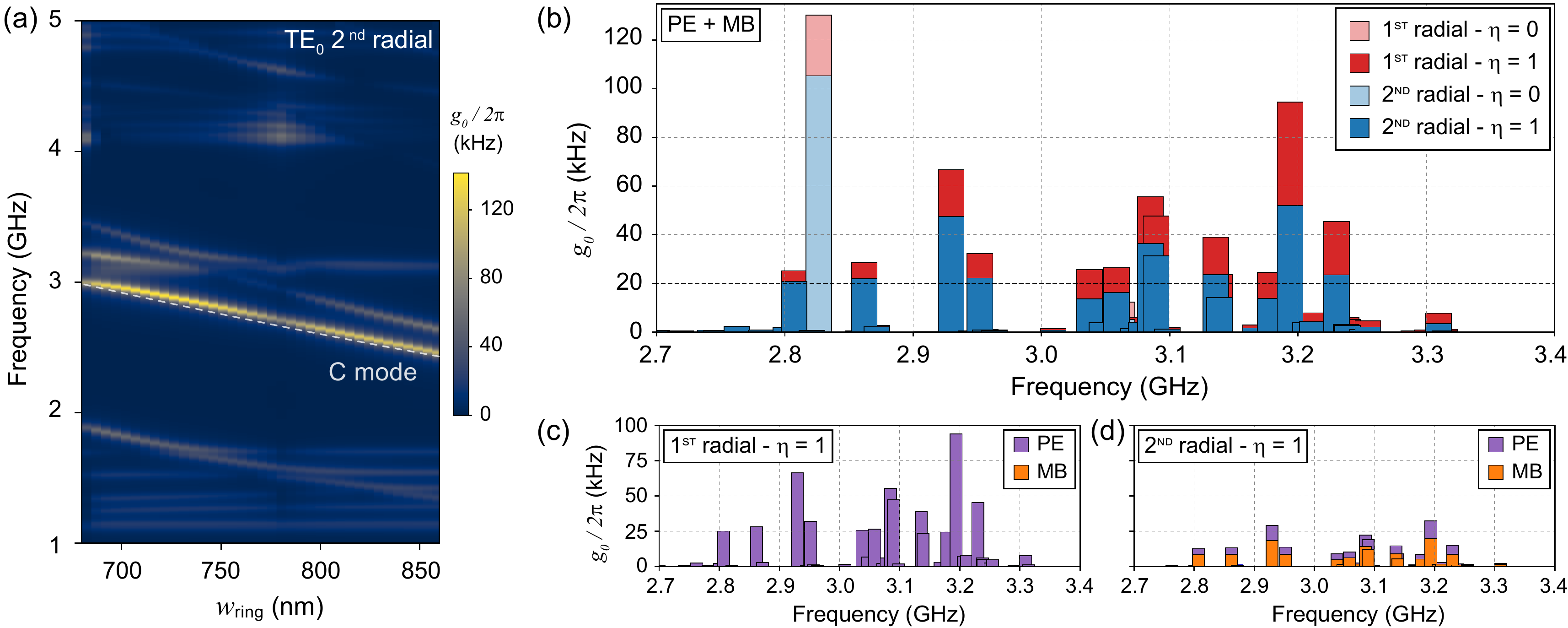}
\caption{(a) 2D FEM results (nominal dimensions) of the optomechanical coupling rate ($g_0/(2\pi)$). The mechanical C-mode structure (discussed in the main text) is highlighted with a gray dashed line as a guide to the eyes. (b) Optomechanical coupling rate obtained from the 3D FEM simulations (real dimensions), where $\eta= 0$ and $\eta = 1$ account for isotropic and anisotropic material elasticity, respectively. PE and MB stand for photo-elastic and moving boundaries coupling mechanisms, as shown specifically in (c) and (d) for the first and second radial-order modes (assuming anisotropic GaAs elasticity). \label{s3}}
\end{figure*}

The comparison of the FEM simulation with the experimental data provided us an unambiguous identification of the measured optical mode, which is Transverse Electric (TE) of second radial-order. Notice that this mode has an avoided crossing with the mode in the first internal ring which explains its resulting anomalous dispersion. Fig. \ref{s3} (a) contains the updated 2D simulation of the optomechanical coupling rate ($g_0/(2\pi)$) for this optical mode structure (using the bullseye nominal dimensions), revealing values very close to those predicted for the fundamental TE optical mode. The coupling to the second-order TE mode, however, only appears when the $w_{\text{ring}}$ approaches \SI{700}{\nm}. 

Finally, the geometrical parameters and optical mode obtained from this analysis were employed in the 3D anisotropic model used in the main text to estimate $g_0$, which agreed well with the measured values. For the sake of completeness, we show in Fig. \ref{s3} (b) the results for the fundamental and second radial-order optical modes considering the real dimensions. There, $\eta = 0$ ($\eta = 1$) stands for isotropic (anisotropic) material elasticity. Furthermore, Figs. \ref{s3} (c) and (d) separate the photo-elastic (PE) and moving boundaries (MB) contributions to the optomechanical coupling rate for both optical modes when $\eta = 1$.

\section{Mechanical linewidth model \label{S3}}

The anharmonic damping experienced by our resonator was computed using~\cite{Vacher2005}:
\begin{equation} \label{eq2}
	\Gamma_{anh} = \omega_m \frac{c_p(T)~T~[\Delta \gamma(T)]^2}{\rho~[c_s(T)]^2} \frac{\omega_m\tau_{ph}(T)}{1+[\omega_m\tau_{ph}(T)]^2},
\end{equation}	

\noindent where $\rho$ is the GaAs mass density and $c_p$~\cite{Blakemore1982}, $(\Delta \gamma)^2$~\cite{Romero1978} and $c_s$~\cite{Cottam1973}  are the volumetric heat capacity, the variance of the Gr\"{u}neisen parameter and the mean Debye sound velocity, whose temperature (T) dependence are considered. As in the main text, $\omega_m$ and $\tau_m$ are the mechanical resonator angular frequency and the anharmonic phonon relaxation time, respectively. 

The attenuation from phonon coupling to two-level systems (TLS) was also obtained from~\cite{Vacher2005} with the ensemble of defects described by a distribution of double-well potentials as in Ref.~\cite{Phillips1987}. Thus, 

\begin{equation} \label{eq3}
\Gamma_{TLS} = \small \frac{\mathcal{C}}{V_1}\Phi \bigg(\frac{\sqrt2 V_1}{\alpha\Delta_1}\bigg)
{\int_0^\infty\alpha x^{- \varepsilon}e^{-\frac{1}{2}x^2}\frac{\omega_m\tau_0~e^{\alpha x}}{1+(\omega_m\tau_0~e^{\alpha x})^2} dx}
\end{equation}	

Here, $\Phi$ is the error function, $\mathcal{C} = \omega_m\vartheta^2N_{TLS}/(\rho~[c_s(T)]^2)$, $\alpha = V_1/ k_B T$ and $k_B$ is the Boltzmann's constant. The parameters $\Delta_1$, $V_1$, $\vartheta$, $N_{TLS}$ and $\tau_0$ represent the double-well potentials asymmetry, the barrier height that separate them, the deformation potential, the density of TLS and the related relaxation time, respectively; all these values were estimated in Ref.~\cite{Hamoumi2018} for a GaAs disk resonator of radius of $\SI{5.5}{\mu m}$ and \SI{200}{nm} thick.

%